\documentclass[12pt]{article}
\usepackage{amsmath,amssymb,graphicx,color,lscape}
\usepackage[square,comma,sort&compress]{natbib}

\newcommand{\al}{\alpha}
\newcommand{\la}{\lambda}

\begin{document}

\begin{titlepage}

\begin{flushright}
MZ-TH/08-26\\[0.2cm]
\today
\end{flushright}

\vspace{0.7cm}
\begin{center}
\Large\bf
On the Perturbative Approach in the\\
Randall-Sundrum Model 
\end{center}

\vspace{0.2cm}
\begin{center}
{\sc Florian Goertz and Torsten Pfoh}\\
\vspace{0.4cm}
{\sl Institut f\"ur Physik (THEP), Johannes Gutenberg-Universit\"at\\
D-55099 Mainz, Germany}
\end{center}

\vspace{0.2cm}
\begin{abstract}\noindent
We study a $SU(2)_L \times U(1)_Y$ gauge theory in the Randall-Sundrum background, including electroweak symmetry breaking due to a brane-localized Higgs sector. We work in the decomposed four dimensional theory and treat the symmetry breaking effects as a perturbation. Although there is an exact approach, this method is widely used and quite intuitive. We extend existing studies by giving explicit analytical expressions in a covariant $R_\xi$ gauge, needed for loop calculations in the decomposed theory. The perturbative approach is also applied to fermions and a comparison of the gauge boson and fermion spectra with the exact results is presented, validating the perturbative method used in the literature.
\end{abstract}
\vfil

\end{titlepage}

\section{Introduction}
 
The Standard Model (SM) of particle physics  has so far resisted every experimental effort to refute its validity. Despite its tremendous success, the mere facts that its interactions do not incorporate gravity and that its gauge couplings do not unify at the grand unification scale, lead to the strong belief that the SM is just a low-energy approximation of a more fundamental theory, valid up to some cutoff.
With no new physics at the TeV scale, the SM interpreted as an effective theory further suffers from the hierarchy problem, namely the issue of how to create and maintain the large separation between the electroweak and the Planck scale, $M_{Pl}$, or some other high scale where new physics sets in.
An elegant solution to the hierarchy problem was proposed by Randall and Sundrum (RS) \citep{Randall:1999ee}. The RS model is formulated in five dimensional (5D) anti de-Sitter space ($AdS_5$) with curvature $k$. The fifth dimension is assumed to be compactified on an orbifold $S^1/Z_2$ with radius $r$ and is labeled by the coordinate $\phi\in[-\pi,\pi]$. Two three-branes, i.e., four dimensional (4D) Minkowskian subspaces, the ultraviolet (UV) and the infrared (IR) brane, are localized at the orbifold fixed points at $\phi=0$ and $\phi=\pi$, respectively. The hierarchy between the electroweak and the Planck scales is then explained by a non-factorizable metric
\begin{equation}
\label{eqn:RSmetric}
ds^2=e^{-2 \sigma(\phi)} \eta_{\mu\nu} dx^\mu dx^\nu - r^2 d\phi^2\, , \quad \sigma(\phi)=k r |\phi|\, ,
\end{equation}
where $\eta_{\mu\nu}=diag(1,-1,-1,-1)$, and $x^\mu$ denote the 4D Minkowski coordinates.
In the original model, the SM fields were assumed to be localized on the IR brane, where mass scales are suppressed by the warp factor $e^{-kr\pi}$, whereas gravity propagates in the bulk, i.e., the whole five dimensional space-time. The parameters $k$ and $1/r$ are taken to be of the order of $M_{Pl}$, with the product $L \equiv k r \pi \approx 37$ chosen in such a way as to account for the ratio of the electroweak and the Planck scales:
\begin{equation}
\epsilon \equiv e^{-L} \approx 10^{-16}.
\end{equation}
It was soon realized that the minimal model could be extended by allowing gauge bosons \citep{Davoudiasl:1999tf,Pomarol:1999ad,Chang:1999nh,Gherghetta:2000qt} and fermions \citep{Grossman:1999ra,Gherghetta:2000qt,Chang:1999nh} to propagate in the bulk, without spoiling the solution to the hierarchy problem. The Higgs field, however, has to live at (or near) the IR brane. Bulk fermions also allow the RS model to address the flavor puzzle, namely admit an explanation for the hierarchical structure of the fermion masses. This is done by localizing the fermions at different positions in the bulk, for which no hierarchical parameters are required \citep{Grossman:1999ra,Gherghetta:2000qt,Huber:2000ie,Huber:2003tu}. 

In this work we study a $SU(2)_L \times U(1)_Y$ gauge theory in the bulk of RS in a covariant $R_{\xi}$ gauge, including mass terms generated by an IR brane-localized Higgs field. We work in the decomposed 4D theory. A formulation in $R_{\xi}$ gauge is important for performing loop calculations, as will be explained in Section \ref{sec:EV}. Within the 5D formulation, $R_{\xi}$ gauges have been studied in \citep{Randall:2001gb,Randall:2001gc}.
There are two different ways of treating the effects of a brane Higgs. The first approach consists of solving the bulk equations of motion (EOM) of the model, without taking into account the Higgs couplings. These are afterwards treated as a perturbation \citep{Pomarol:1999ad,Grossman:1999ra,Gherghetta:2000qt,delAguila:2000kb,Huber:2000fh,Hewett:2002fe,Huber:2003tu}. The other approach includes the Higgs-induced mass terms from the beginning, which enter the EOM through boundary conditions (BCs) \citep{Gherghetta:2000qt,Csaki:2003sh,Bagger:2001qi,Casagrande:2008hr}. We will refer to this as the exact approach. As the first approach is widely used in the literature and also gives an intuitive explanation of mixings between the modes, we will employ this technique in the following. A central part of this work is a comparison of the spectra that one obtains by using the different methods, which provides a cross-check of the results. Finally, we extend the study to bulk fermions. 

Our work is organized as follows.
In Section \ref{sec:bulkgauge}, the electroweak gauge theory in the bulk of RS is formulated in  $R_{\xi}$ gauge, using a 4D description.\footnote{It is straightforward to extend the results to the $SU(3)_C$ gauge group.} The mass matrices are diagonalized analytically, and formulas for the masses and mixings are given. In addition a numerical analysis of the spectrum is performed, comparing the results with the findings of the exact approach. In Section \ref{sec:fermions}, we consider a single bulk fermion and also diagonalize the mass matrix analytically. We then perform similar numerical studies as in the case of the gauge boson sector. Our conclusions are presented in Section \ref{sec:Conclusions}.

\section{Bulk Gauge Fields in $R_\xi$ Gauge}
\label{sec:bulkgauge}

In this section we derive the 4D theory for the electroweak gauge sector in a covariant $R_\xi$ gauge. The IR brane-localized Higgs field is introduced as a perturbation which is coupled to the (unperturbed) states after Kaluza-Klein (KK) decomposition. After the fifth dimension is integrated out, the resulting mass matrices for the vector and scalar fields have to be diagonalized in order to obtain the mass eigenstates. Due to their regular structure this is possible, despite the fact that the mass matrices are infinite dimensional. We will find that the Goldstone bosons of the Higgs doublet mix with the scalar components of the decomposed 5D gauge fields. We split up the 5D action as in \citep{Casagrande:2008hr}
\begin{equation}
S_{gauge}=\int d^4x\, r \int_{-\pi}^\pi d\phi \left({\cal L}_{W,B}+{\cal L}_{Higgs}+{\cal L}_{GF}\right) + {\rm ghost\ terms}\, ,
\label{eqn:action}
\end{equation}
but postpone the rotations from $A^a=(W^1,W^2,W^3,B)^T$ to the physical states $(W^\pm,Z,A)$.
Thus, the bilinear terms resulting from the covariant derivative of the Higgs doublet read
\begin{equation}
\begin{split}
{\cal L}_{Higgs} \ni\, & \frac{\delta(|\phi|-\pi)}{r}\ (D_\mu \Phi)^\dagger (D^\mu \Phi)\Big{|}_{\mbox{\tiny 2}} \\
=\, & \frac{\delta(|\phi|-\pi)}{2\, r}\ \left[(\partial_\mu h)^2+(\partial_\mu \varphi_i)^2 - 2 g_5^{(a)}F^a_{\ i}\, \varphi_i\partial^\mu A_\mu^a + g_5^{(a)}g_5^{(b)}F^a_{\ i}F^b_{\ i}A_\mu^a A^{\mu b}\right],
\label{eqn:higgskinetic}
\end{split}
\end{equation}
containing the $a\times i$ matrix \citep{PS}
\begin{equation}
\left\{g_5^{(a)} F^a_{\ i}\right\}=\frac v2 
\left(\begin{array}{*{3}{c}}
g_5\ & 0 & 0 \\
0\ \ & g_5 & 0 \\
0\ \ & 0 & g_5 \\
0\ \ & 0 & -g_5'\\
\end{array}\right)\, . 
\label{eqn:F}
\end{equation}
The index $i$ counts degrees of freedom perpendicular to the Higgs field, whose vacuum expectation value (VEV) is denoted by $v \approx 246$\, GeV, and summation over $a, b$, and $i$ is understood.
The bilinear terms in the pure gauge sector read
\begin{equation}
{\cal L}_{W,B,\, 2}=-\frac 14 F_{\mu\nu}^aF^{\mu\nu a}
+ \frac {e^{-2\sigma}}{2r^2}\Big{(}\partial_\phi A_\mu^a \partial_\phi A^{\mu a} + \partial_\mu A_5^a \partial^\mu A_5^a\Big{)} - \partial^\mu A_\mu^a\partial_\phi \frac {e^{-2\sigma}}{r^2} A_5^a \, .
\label{eqn:gaugekinetic}
\end{equation}
The mixing terms in (\ref{eqn:higgskinetic}) and (\ref{eqn:gaugekinetic}) could be removed by choosing the following gauge fixing
\begin{equation}
{\cal L}_{GF}=- \frac{1}{2\xi_a} \left[\partial^\mu A_\mu^a-\xi_a\left(\frac {\delta(|\phi|-\pi)}r\, g_5^{(a)} F^a_{\ i}\varphi_i + \partial_\phi\frac{e^{-2\sigma(\phi)}}{r^2} A_5^a\right)\right]^2\, .
\label{eqn:GF5D}
\end{equation}

\subsection{Kaluza-Klein Decomposition}

The gauge fixing Lagrangian (\ref{eqn:GF5D}) contains terms in which the $\delta$-distributions get squared. In \citep{Casagrande:2008hr} it was shown, that these terms cancel with $\delta$-contributions from derivatives of the scalar components of the gauge fields, entering through the use of the exact EOM. As we work with unperturbed fields, the relevant terms in the EOM are absent and we would have to find a method to deal with the $\delta^2$-terms, e.g. inserting the completeness relation for one $\delta$ after KK decomposition. Alternatively, performing the integral over the fifth dimension {\it before} fixing the gauge removes the $\delta$-distributions from the theory and we will proceed in this way. Therefore we first perform the KK decomposition and write
\begin{equation}
S_{gauge}=\int d^4x \left[r \int_{-\pi}^\pi d\phi \left({\cal L}_{W,B}+{\cal L}_{Higgs}\right)+{\cal L}_{GF}^{4D}\right]+{\rm ghost\ terms}\, ,
\label{eqn:actionmod}
\end{equation}
where we have introduced a purely 4D gauge fixing Lagrangian.
The KK decompositions for the gauge fields read \citep{Davoudiasl:1999tf,Randall:2001gb}
\begin{equation}
\begin{split}
A_\mu^a(x,\phi)&=\frac 1{\sqrt r}\sum_n A_\mu^{a(n)}(x) \chi_n(\phi), \\
A_5^a(x,\phi)&=- \frac 1{\sqrt r}\sum_n A_5^{a(n)}(x) \frac 1{m_n}\partial_\phi\chi_n(\phi)\, .
\end{split}
\end{equation}
The profiles $\chi_n$ are universal for all gauge fields and form complete sets of even functions on the orbifold, obeying orthonormality conditions and the well known EOM derived in \citep{Davoudiasl:1999tf}.
As the decomposition of the scalar modes contains the $Z_2$-parity odd term $\partial_\phi\chi_n$, there is no zero mode. Thus, the summation starts at $n=1$ for that case, whereas it runs from zero to infinity for the vector fields. This convention will be used throughout the paper.

We define the dimensionless coupling constant $g^{(a)}=g^{(a)}_5/\sqrt{2\pi r}$ and remove the mixing terms in the decomposed 4D action by introducing the gauge fixing Lagrangian
\begin{equation}
{\cal L}_{GF}^{4D}=-\frac 1{2\xi_a}\sum_n \left[\partial^\mu A_\mu^{a(n)}-\xi_a\left(\sqrt{2\pi}g^{(a)} F^a_{\ i}\varphi_i\chi_n(\pi)+m_n A_5^{a(n)}\right)\right]^2\equiv -\frac 12\sum_n {G^{a(n)}}^2\, .
\label{eqn:GF4D}
\end{equation}
Thus, making use of the EOM and integrating over $\phi$, the bilinear terms in the Lagrangian finally take the form
\begin{equation}
\label{eqn:L2}
\begin{split}
{\cal L}^{4D}_{gauge,2} =&\sum_n\Big{[}-\frac 14 F_{\mu\nu}^{a(n)}F^{\mu\nu a(n)} - \frac 1{2\xi_a}(\partial^\mu A_\mu^{a(n)})^2 + \frac 12 \sum_m (m^2)_{mn}^{ab} A_\mu^{a(m)} A^{\mu b(n)} \\
& + \frac 12 (\partial_\mu h)^2-\la v^2 h^2 + \frac 12 (\partial_\mu \varphi_i)^2 + \frac 12 \partial_\mu A_5^{a(n)}\partial^\mu A_5^{a(n)} \\
& - \frac {\xi_a}2 \Big{(}(m^2)_{nn}^{ij}\varphi_i\varphi_j + 2\sqrt{2\pi}\ m_n \chi_n(\pi)g^{(a)}F^a_{\ i}\varphi_i A_5^{a(n)}+m_n^2 {A_5^{a(n)}}^2\Big{)}\Big{]}\\ & + {\rm ghost\ terms}\, .
\end{split}
\end{equation}
Here, we have introduced
\begin{align}
(m^2)_{mn}^{ab}=&m_n^2 \delta_{mn}\delta_{ab} + 2\pi\ \chi_m(\pi)\chi_n(\pi) \ g^{(a)}g^{(b)} F^a_{\ i}F^b_{\ i}\, ,\label{eqn:masses}\\
(m^2)_{nn}^{ij}=&2\pi\ \chi_n(\pi)^2\ {g^{(a)}}^2 F^a_{\ i}F^a_{\ j}\, ,
\end{align}
containing the expressions \citep{PS}
\begin{equation}
\begin{split}
g^{(a)}g^{(b)} F^a_{\ i}F^b_{\ i}=g^{(a)}g^{(b)} (FF^T)^{ab}=\frac {v^2}4
\left(\begin{array}{*{4}{c}}
g^2 & 0 & 0 & 0 \\
0 & g^2 & 0 & 0 \\
0 & 0 & g^2 & -g g^\prime \\ 
0 & 0 & -g g^\prime & {g^\prime}^2 \\
\end{array}\right),\\
{g^{(a)}}^2 F^a_{\ i}F^a_{\ j}={g^{(a)}}^2 \Big{(}F^TF\Big{)}_{ij}=\frac {v^2}4
\left(\begin{array}{*{3}{c}}
g^2\ \ & 0 & 0 \\
0\ \ & g^2 & 0 \\
0\ \ & 0 & g^2+{g^\prime}^2 \\
\end{array}\right),
\end{split}
\end{equation}
well known from the SM.

\subsection{Diagonalization of the Mass Matrices}

Now one has to perform two types of diagonalizations. The first one concerns the two matrices shown above and is realized by applying the usual field redefinitions
\begin{align}
&W_{\mu,5}^{(n)\pm}=\frac 1{\sqrt 2}(W_{\mu,5}^{(n)1}\mp iW_{\mu,5}^{(n)2})\, ,\quad \varphi^\pm=\frac 1{\sqrt 2}(\varphi_1\mp i\varphi_2)\, ,\nonumber\\
&Z_{\mu,5}^{(n)}=\frac 1{\sqrt{g^2+{g^\prime }^2}}(g W_{\mu,5}^{(n)3} -g^\prime B_{\mu,5}^{(n)})\ ,\label{eqn:basistrafo}\\
&A_{\mu,5}^{(n)}=\frac 1{\sqrt{g^2+{g^\prime }^2}}(g^\prime W_{\mu,5}^{(n)3} +g B_{\mu,5}^{(n)})\ ,\nonumber
\label{eqn:redefi}
\end{align}
and $\varphi^0=\varphi_3$.
Thus, the mixing term for the scalars in  (\ref{eqn:L2}) takes the form
\begin{equation}
g^{(a)}F^a_{\ i}\varphi_i A_5^{a(n)}=\frac v2 \left(g\,(\varphi^+ W_5^{(n)-}+\varphi^- W_5^{(n)+})+\sqrt{g^2+{g^\prime}^2}\,\varphi^0 Z_5^{(n)}\right)
\end{equation}
and the whole Lagrangian decomposes into separate Lagrangians for the fields $W^{\pm(n)}$, $Z^{(n)}$, and $A^{(n)}$.

The second type of diagonalization concerns the mixings of the different KK modes and was discussed in detail for the case of the SM with a flat extra dimension in \citep{Muck:2001yv}, using a technique first presented in \citep{Dienes:1998sb}. A glimpse on the $\xi$-dependent mass term in (\ref{eqn:L2}) tells us that the Goldstone bosons mix with the KK tower of the corresponding scalar components of the gauge fields. Therefore we introduce the infinite dimensional vectors
\begin{align}
&W^\pm_5=(\varphi^\pm, W^{\pm(1)}_5, W^{\pm(2)}_5,\ ...)^T\, ,\nonumber\\
&Z_5=(\varphi^0, Z^{(1)}_5, Z^{(2)}_5,\ ...)^T\, ,\\
&A_5=(0\ , A^{(1)}_5, A^{(2)}_5,\ ...)^T.\nonumber 
\end{align}
The mass terms of the scalar fields in (\ref{eqn:L2}) then take the form
\begin{equation}
{\cal L}^\xi_{mass}=-\xi_W W_5^{+T} {M_W^\xi}^2 W_5^- -\frac {\xi_Z} 2 Z_5^T {M_Z^\xi}^2 Z_5 -\frac {\xi_A} 2 A_5^T {M_A^\xi}^2 A_5\, .
\end{equation}
The squared mass matrix
\begin{equation}
{M_X^{\xi}}^2=
\left(\begin{array}{*{5}{c}}
\sum_{n=0}{m_X^{(n,n)}}^2 & m_X^{(1,1)}m_1 & m_X^{(2,2)}m_2 & m_X^{(3,3)}m_3 & \cdots \\
m_X^{(1,1)}m_1 & m_1^2 & 0 & 0 & \cdots \\
m_X^{(2,2)}m_2 & 0 & m_2^2 & 0 & \cdots \\ 
m_X^{(3,3)}m_3 & 0 & 0 & m_3^2 & \cdots \\
\vdots & \vdots & \vdots & \vdots & \ddots \\
\end{array}\right)
\label{eqn:scalarmatrix}
\end{equation}
contains the bare KK masses $m_n$ (which are equal for all types of gauge bosons) as well as individual mass terms $m^{(m,n)}_X$, $X=W,Z$ that arise from electroweak symmetry breaking (EWSB) and are given by
\begin{equation}
\begin{split}
{(m^{(m,n)}_W)}^2&=2\pi\ \frac{g^2 v^2}4 \chi_m(\pi)\chi_n(\pi)\ ,\\
{(m^{(m,n)}_Z)}^2&=2\pi\ \frac {(g^2+{g^\prime }^2) v^2}4 \chi_m(\pi)\chi_n(\pi)\, .
\label{eqn:inmass}
\end{split}
\end{equation}
One might worry about the divergent $(0,0)$ component in (\ref{eqn:scalarmatrix}), which is related to the squared $\delta$-distribution in the 5D gauge fixing Lagrangian (\ref{eqn:GF5D}). In fact, this divergence vanishes after diagonalization as we will see below. After implementation of the basis transformation (\ref{eqn:basistrafo}) and introducing the vector $W^\pm_\mu=(W^{\pm(0)}_\mu, W^{\pm(1)}_\mu, W^{\pm(2)}_\mu,\ ...)^T$ and similarly for $Z_\mu$ and $A_\mu$, the mass matrix for the gauge fields (\ref{eqn:masses}) can also be expressed in terms of the above definitions~(\ref{eqn:inmass}). It reads \citep{Hewett:2002fe}
\begin{equation}
{M_X}^2=
\left(\begin{array}{*{5}{c}}
{m_X^{(0,0)}}^2 & {m_X^{(0,1)}}^2 & {m_X^{(0,2)}}^2 & {m_X^{(0,3)}}^2 & \cdots \\
{m_X^{(1,0)}}^2 & m_1^2+{m_X^{(1,1)}}^2 & {m_X^{(1,2)}}^2 & {m_X^{(1,3)}}^2 & \cdots \\
{m_X^{(2,0)}}^2 & {m_X^{(2,1)}}^2 & m_2^2+{m_X^{(2,2)}}^2 & {m_X^{(2,3)}}^2 & \cdots \\ 
{m_X^{(3,0)}}^2 & {m_X^{(3,1)}}^2 & {m_X^{(3,2)}}^2 & m_3^2+{m_X^{(3,3)}}^2 & \cdots \\
\vdots & \vdots & \vdots & \vdots & \ddots \\
\end{array}\right)
\label{eqn:vectormatrix}
\end{equation}
and the mass term becomes
\begin{equation}
{\cal L}_{mass}=W_\mu^{+T} {M_W}^2 W^{-\mu} +\frac 12 Z_\mu^T {M_Z}^2 Z^\mu +\frac 12 A_\mu^T {M_A}^2 A^\mu\, .
\end{equation}
Note that for the photon, the mass matrices (\ref{eqn:scalarmatrix}) and (\ref{eqn:vectormatrix}) reduce to the same diagonal matrix, containing the KK masses $m_n$ and possessing a vanishing $(0,0)$ component. For the $W$ and $Z$ bosons, additional work is required, but it will turn out, that the two matrices indeed have the {\it same} eigenvalues as needed to maintain gauge invariance. Defining
\begin{equation}
m_W=\frac {g v}2\, ,\quad m_Z=\frac {\sqrt{g^2+{g^\prime }^2}\ v}2\, ,\quad\al_n=\sqrt{2\pi}\chi_n(\pi),
\end{equation}
where $\al_0=1$ (see below), we derive from (\ref{eqn:scalarmatrix}) and (\ref{eqn:vectormatrix}) the characteristic polynomials
\begin{align}
\det({M_X^\xi}^2-\la{\textbf 1})&=m_X^2\Big{[}\Big{(}1+\sum_{n=1}\al_n^2-\frac \la{m_X^2}\Big{)}\prod_{n=1}(m_n^2-\la)-\sum_{n=1}\al_n^2m_n^2\prod_{k\not =n}(m_k^2-\la)\Big{]}\, ,\\
\det(M_X^2-\la{\textbf 1})&=\prod_{n=0}(m_n^2-\la)+m_X^2\sum_{n=0}\al_n^2\prod_{k\not =n}(m_k^2-\la)\ .
\end{align}
After some algebraic manipulation, {\it both} equations take the form
\begin{equation}
\det({M_X^{(\xi)}}^2-\la{\textbf 1})=\Big{(}\prod_{n=1}(m_n^2-\la)\Big{)}\Big{(}m_X^2-\la-\la\, m_X^2\sum_{n=1}\frac {\al_n^2}{m_n^2-\la}\Big{)}.
\end{equation}
Since $v \not = 0$ implies $m_n^2 \not = \la$, the squared mass eigenvalues $\la$ are given by the transcendental equation 
\begin{equation}
m_X^2-\la-\la\, m_X^2\sum_{n=1}\frac {\al_n^2}{m_n^2-\la}=0,
\label{eqn:trans}
\end{equation}
which generalizes the result from \citep{Muck:2001yv} to the case of a warped extra dimension. We stress that, although the Higgs has been introduced as a perturbation, the latter equation is an exact result, as long as one does not truncate the sum. In the following we use the notation $ M_X^{(n)2} \equiv \la_n$, for the $n$-th solution to (\ref{eqn:trans}).

\subsection{The Spectrum}

As equation (\ref{eqn:trans}) can not be solved analytically, we perform an iterative solution in powers of the ratio $v^2/M_{KK}^2$, where $M_{KK} \equiv k \epsilon = k e^{-k r \pi}$ is the mass scale of the low-lying KK~excitations. This leads to
\begin{equation}
\label{eqn:boexp}
\begin{split}
M_X^{(0)2}=&m_X^2 \left(1-m_X^2\sum_{n=1}\frac {\al_n^2}{m_n^2}+\mathcal{O}\left(\frac{m_X^4}{M_{KK}^4}\right) \right),\\
M_X^{(n)2}=&m_n^2\left(1+\frac{m_X^2}{m_n^2}\al_n^2+\frac{m_X^4}{m_n^4}\al_n^2\Big{(}1+\sum_{k\not =n}\frac {m_n^2 \al_k^2}{m_n^2-m_k^2}\Big{)}+ \mathcal{O}\left(\frac{m_X^6}{M_{KK}^6}\right) \right)\, .
\end{split}
\end{equation} 
One observes that the mass of the zero mode decreases, compared to the bare value $m_X^2$, whereas the masses of the excitations increase compared to the unperturbed case. Furthermore, the mass of the scalar zero mode is finite. 

In order to compare the masses numerically with the results of \citep{Casagrande:2008hr}, we need the solutions to the bulk EOM for the gauge fields, entering the above equations. They are given in \citep{Davoudiasl:1999tf,Gherghetta:2000qt,Casagrande:2008hr}.
Introducing the variables $t=\epsilon e^\sigma(\phi) \in [\epsilon,1]$ \citep{Grossman:1999ra} and $x_n=m_n/M_{KK}$, they read
\begin{equation}
\chi_n(\phi)=N_n\sqrt{\frac L\pi}\, t\, c_n^+(t)\, ,
\end{equation}
with
\begin{equation}
\begin{split}
c_n^+(t)&=Y_0(x_n\epsilon)J_1(x_n t)-J_0(x_n\epsilon)Y_1(x_n t)\, ,\\
N_n^{-2}&=[c_n^+(1)]^2
-\epsilon^2[c_n^+(\epsilon)]^2\, ,
\end{split}
\end{equation}
for the case of a vanishing Higgs coupling.
The masses $m_n$ of the unperturbed states are derived from the BC $\partial_\phi \chi_n|_{\phi=\pi}=0$, which translates into
\begin{equation}
        c_n^-(1) \equiv Y_0(x_n\epsilon)\,J_0(x_n) - J_0(x_n\epsilon)\,Y_0(x_n)=0,
\end{equation}
and there exists also a constant zero mode solution $\chi_0(\phi)=\frac{1}{\sqrt{2 \pi}}$, with $m_0=0$.
In the exact approach, the masses of the physical states are derived from the condition \citep{Casagrande:2008hr}
\begin{equation}
\label{eqn:exbos}
   x_n\,c_n^-(1) 
   = - \frac{g^2 v^2}{4 M_{KK}^2}\,L\,c_n^+(1).
\end{equation}
\begin{figure}[t!]
        \begin{center}
  \includegraphics[width=7cm]{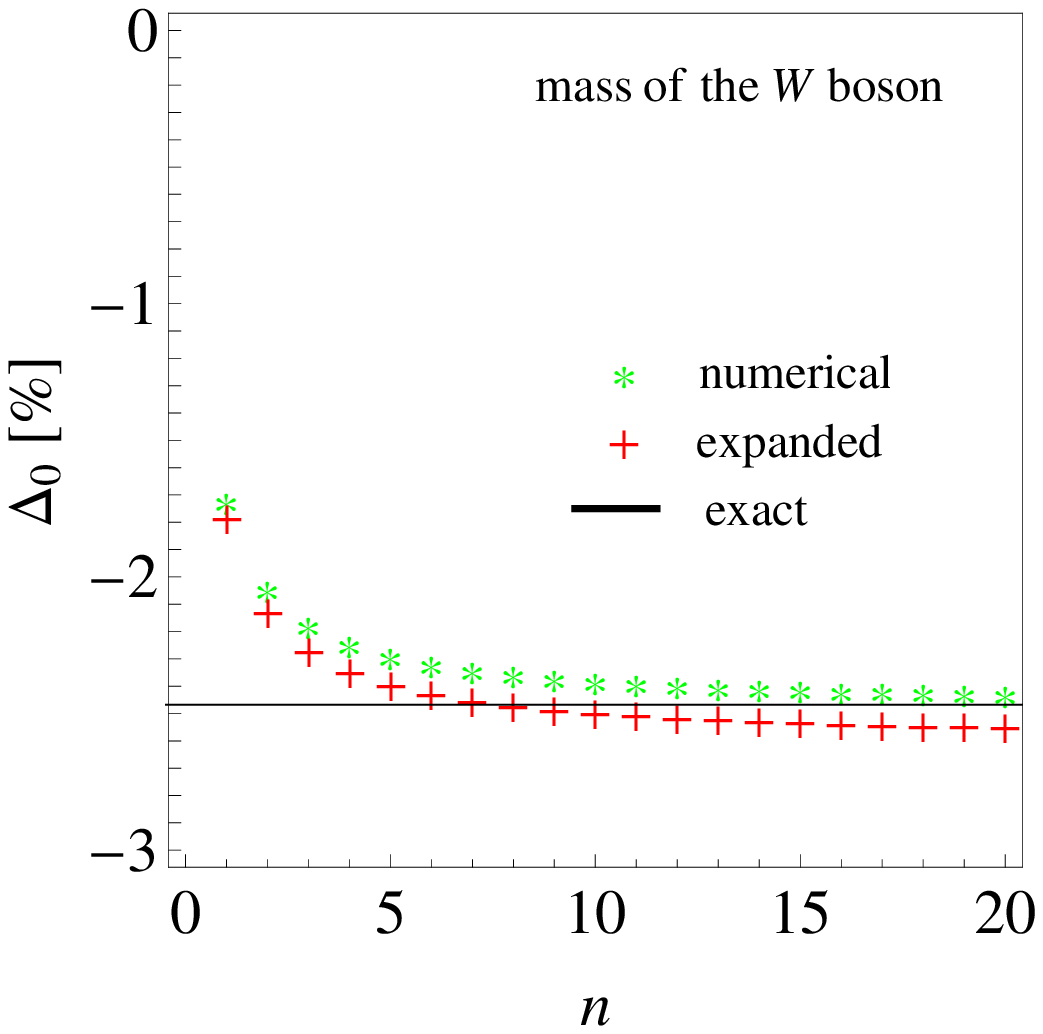}\qquad \includegraphics[width=7.35cm]{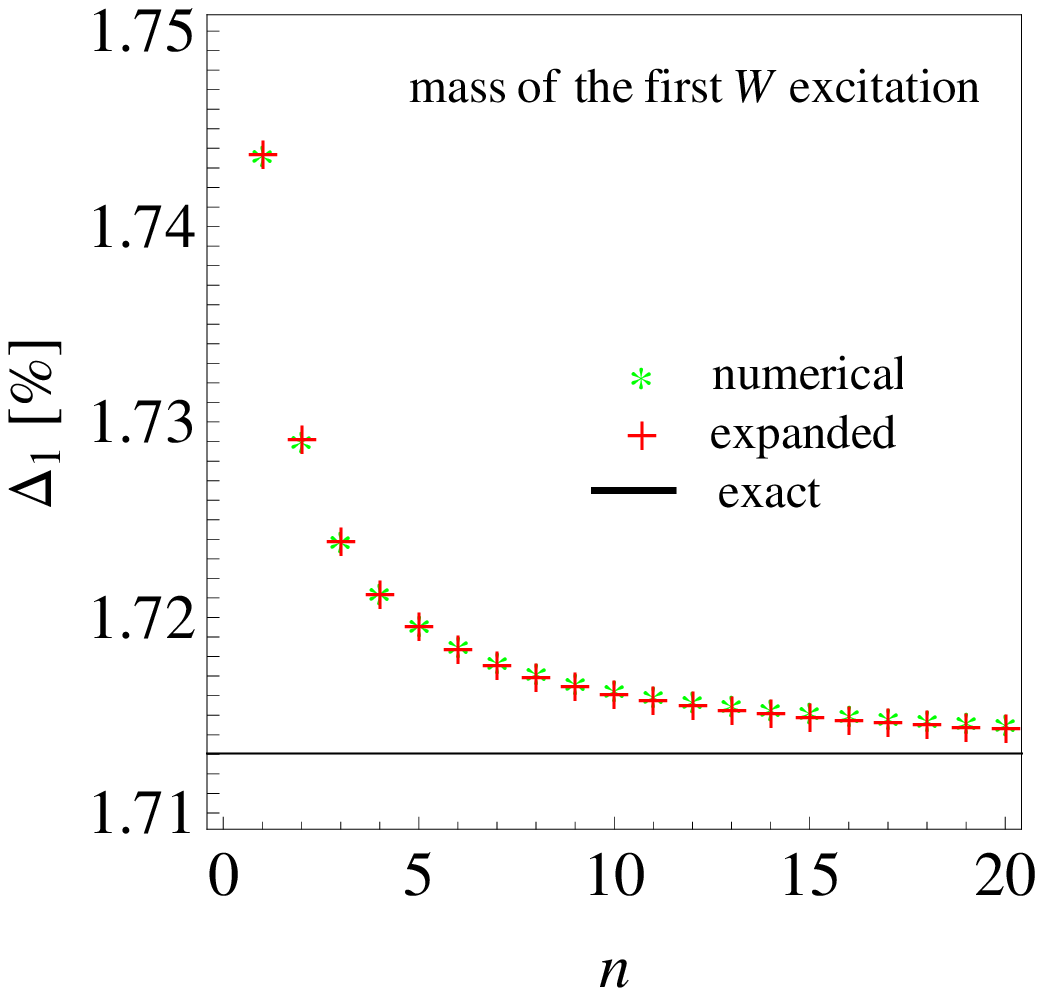}
        \end{center}
        \caption{Dependence of the masses of the $W$-boson and its first excitation on the truncation.}´
        \label{fig:gaugecon}
\end{figure}

After truncation we can solve the transcendental equation (\ref{eqn:trans}) numerically. In Figure \ref{fig:gaugecon} we compare the masses of the $W$ boson and its first excitation that we get by the different approaches of treating the Higgs, illustrating the dependence on the truncation. Here and in the following we use $\epsilon=10^{-16}$ and $M_{KK}=1.5$\, TeV as input parameters. We plot relative deviations of the perturbed masses from the bare ones, denoted by $\Delta_i \equiv M_W^{(i)}/m_i -1$, where $m_0=m_W$. The green (light gray) stars correspond to a numerical solution of (\ref{eqn:trans}), truncating the infinite sum after $n$ modes. The red (dark gray) crosses visualize the masses that we get from the expansion up to $\mathcal{O}(m_X^4/M_{KK}^2)$ (see (\ref{eqn:boexp})), truncated in the same way, and the straight black lines represent the exact results obtained by using the formalism of \citep{Casagrande:2008hr} (see (\ref{eqn:exbos})). One observes that the perturbative approach works very well for the masses of the gauge bosons. The numerical solutions converge quickly to the exact values. It is noteworthy that taking into account only the mixing with the first KK level, i.e., truncating the sum in (\ref{eqn:trans}) at $n=1$, already gives about $ 70 \%$ of the total correction to the zero mode mass. For the zero mode the limit of the expansion lies slightly below the true value due to the neglected $\mathcal{O}(m_X^6/M_{KK}^4)$ contributions. However, already the first corrections in the expansion (\ref{eqn:boexp}) lead to results very close to the exact value and it's a good approximation to truncate the sum at low $n$.

\subsection{Eigenvectors}
\label{sec:EV}
In the following we give the explicit form of the diagonalization matrices for the massive gauge bosons $X=W,Z$, defined by 

\begin{equation}
\begin{split}
&B^T {M_X}^2 B= \tilde M_X^2\, ,\quad BB^T={\bf 1}\, ,\\
&G^T {M_X^\xi}^2 G = \tilde M_X^2,\quad GG^T={\bf 1}\, ,
\end{split}
\end{equation}
where $ {\tilde M_X}^2$ denotes the diagonal matrix built out of the solutions $\la_n$ to (\ref{eqn:trans}). We start with the diagonalization for the 4D vector fields. Calculating the eigenvectors of ${M_X}^2$, we get the relation between mass eigenstates $\tilde{X}_\mu^{(n)}$ and interaction eigenstates $X_\mu^{(n)}$: 
\begin{equation}
\tilde{X}_\mu^{(n)}=\sum_m B^{T(n,m)}{X}_\mu^{(m)}=\frac{1}{\sqrt{1+\sum_{j=1}\frac{\al_j^2 \la_n^2}{(m_j^2-\la_n)^2}}}\left(X_\mu^{(0)}-\sum_{m=1} \frac{\al_m \la_n}{m_m^2-\la_n} X_\mu^{(m)} \right).  
\label{eqn:BMatrix}
\end{equation}
In the same way we obtain
\begin{equation}
\tilde{X_5}^{(n)}=\sum_m G^{T(n,m)}{X_5}^{(m)}=\frac{1}{\sqrt{1+\sum_{j=1}\frac{\al_j^2 m_j^2 m_X^2}{(m_j^2-\la_n)^2}}} \left(X_5^{(0)}-\sum_{m=1} \frac{\al_m m_m m_X}{m_m^2-\la_n} X_5^{(m)} \right)
\label{eqn:GMatrix}     
\end{equation} 
for the scalars.
To get a better feeling for the structure of the mixing matrices, we plug in the expansions for the mass eigenvalues (\ref{eqn:boexp}) up to $\mathcal{O}(m_X^2/M_{KK}^2)$ and arrive at:
\begin{equation}
\begin{split}
B=& \left(
\begin{array}{ccccc}
        1 & \al_1 \frac{m_X^2}{m_1^2} & \al_2 \frac{m_X^2}{m_2^2} &  \al_3 \frac{m_X^2}{m_3^2} & \dots \\
        - \al_1 \frac{m_X^2}{m_1^2} & 1 & \al_2 \al_1 \frac{m_X^2}{m_2^2-m_1^2} & \al_3 \al_1 \frac{m_X^2}{m_3^2-m_1^2}
 & \dots \\
- \al_2 \frac{m_X^2}{m_2^2} & - \al_1 \al_2  \frac{m_X^2}{m_2^2-m_1^2} & 1 &  \al_3 \al_2 \frac{m_X^2}{m_3^2-m_2^2} & \dots \\ 
- \al_3 \frac{m_X^2}{m_3^2} & - \al_1 \al_3 \frac{m_X^2}{m_3^2-m_1^2} & - \al_2 \al_3 \frac{m_X^2}{m_3^2-m_2^2} &  1 & \dots \\ 
\vdots & \vdots & \vdots & \vdots & \ddots 
\end{array}\right)\, ,\\
G=& \left(
\begin{array}{ccccc}
        1-\frac 1 2 m_X^2 \sum_{n=1} \frac{\al_n^2}{m_n^2} & \al_1 \frac{m_X}{m_1} & \al_2 \frac{m_X}{m_2} &  \al_3 \frac{m_X}{m_3} & \dots \\
         - \al_1 \frac{m_X}{m_1} & 1 -\frac 1 2 m_X^2 \frac{\al_1^2}{m_1^2}& m_X^2 \frac{m_1}{m_2} \frac{\al_2 \al_1}{m_2^2-m_1^2} &  m_X^2 \frac{m_1}{m_3} \frac{\al_3 \al_1}{m_3^2-m_1^2} 
 & \dots \\
 - \al_2 \frac{m_X}{m_2} & - m_X^2 \frac{m_2}{m_1} \frac{\al_1 \al_2}{m_2^2-m_1^2}  & 1 -\frac 1 2 m_X^2 \frac{\al_2^2}{m_2^2} &  m_X^2 \frac{m_2}{m_3} \frac{\al_3 \al_2}{m_3^2-m_2^2}  & \dots \\ 
 - \al_3 \frac{m_X}{m_3} & - m_X^2 \frac{m_3}{m_1} \frac{\al_1 \al_3}{m_3^2-m_1^2}  & - m_X^2 \frac{m_3}{m_2} \frac{\al_2 \al_3}{m_3^2-m_2^2}  &  1 -\frac 1 2 m_X^2 \frac{\al_3^2}{m_3^2} & \dots \\ 
\vdots & \vdots & \vdots & \vdots & \ddots
\end{array}\right).
\end{split}
\end{equation}
From these expressions one can see that for the scalars, the mixings between the zero mode and the KK excitations is only suppressed by $\mathcal{O}(m_X/M_{KK})$, compared to $\mathcal{O}(m_X^2/M_{KK}^2)$ for the vectors. It is easy to check that the given expansions for $B$ and $G$ diagonalize the corresponding mass matrices exactly to second order in $m_X/M_{KK}$.

The propagators for the massive gauge bosons and corresponding scalars are easily derived from (\ref{eqn:L2}), after rotating to mass eigenstates. They look like the standard propagators for massive gauge and Goldstone bosons. The fact that the same mass $M_X^{(n)2}$ appears in both, is essential for the cancellation of the dependence on the gauge fixing parameter $\xi$ in amplitudes. Indeed, the towers of scalars $\tilde X^{(n)}_5$ play the role of Goldstone bosons which are absorbed into the longitudinal components of the gauge boson towers.
The limit $\xi \rightarrow \infty$ corresponds to the unitary gauge in which the Goldstone bosons are completely removed from the theory. This gauge has often been used in the literature \citep{Davoudiasl:1999tf,Chang:1999nh,Pomarol:1999ad,Huber:2000fh}. However, if one wants to perform loop calculations involving gauge bosons with massive zero modes, one has to be careful. In general the integration over loop momenta does not commute with the limit $\xi \rightarrow \infty$. Thus, removing the Goldstone bosons can lead to problems, when there are several mass scales in the theory, which is the case in the RS model, where one deals with $m_X$ and $M_{KK}$. Therefore it is important to perform the gauge fixing in a covariant $R_\xi$ gauge. It should be mentioned that inner lines can also be expressed through 5D propagators, which has the advantage that one does not have to sum over the KK tower \citep{Randall:2001gb}.

\section{Fermions}
\label{sec:fermions}
In the following we consider bulk fermions, coupled to a brane-localized Higgs sector, and diagonalize the mass matrices in closed form. We restrict ourselves to the case of just one single fermion, as the internal mixings of $N$ generations lead to additional difficulties (see end of Section \ref{sec:fermions}). This simplified scenario illustrates the capabilities of the perturbative approach. The KK decomposition of the action as well as the bulk EOM can be found in \citep{Grossman:1999ra}. Remember that in order to reproduce the SM at low energies, one needs (besides three generations) two sets of fermions, one charged under $SU(2)_L$ ($Q$), with a left-handed zero mode, and one singlet ($q$), with a right-handed zero mode. Since we consider just one single fermion, in the following the labels $Q$ ($q$) denote particles that are charged (neutral) under the corresponding gauge group.
We introduce the Yukawa action
\begin{equation}
        S_{Y}=-\int{d^4x}\int{d\phi}\sqrt{G} 
 \left(\la^{(5)}_{f} \bar\Psi_L^{(Q)} e^\sigma \Phi\Psi_R^{(q)} + h.c. \right) \delta{(\phi-\pi)}\, .
 \label{eqn:yu}
\end{equation}
Note that the convention for the 5D Yukawa coupling above differs from that in \citep{Casagrande:2008hr} by a factor of $r$.
Plugging in the KK decomposition and integrating over $\phi$ leads to the mass term
\begin{align}
\mathcal{L}_{m}=-\sum_{m,n} m_{f}^{(m,n)} \bar\psi_m^{L(Q)}(x) \psi_n^{R(q)}(x) + h.c., 
\end{align} 
with
\begin{equation}
        m_{f}^{(m,n)}=\frac{v}{\sqrt 2}\ \epsilon^{-1}\,\la^{(5)}_{f}\, f_m^{L(Q)}(\pi) f_n^{R(q)}(\pi)\,.
        \label{eqn:Fermmass}
\end{equation}
After EWSB we can combine $\psi_n^{L(Q)}$ with $\psi_n^{L(q)}$ and $\psi_n^{R(Q)}$ with $\psi_n^{R(q)}$ into the vectors 
\begin{equation}
\begin{split}
        \widehat\Psi_L \equiv \left(\psi_0^{L(Q)},\psi_1^{L(Q)},\psi_1^{L(q)},\psi_2^{L(Q)},\psi_2^{L(q)}, \dots \right)^T,\\
        \widehat\Psi_R \equiv \left(\psi_0^{R(q)},\psi_1^{R(Q)},\psi_1^{R(q)},\psi_2^{R(Q)},\psi_2^{R(q)}, \dots \right)^T.
        \end{split}
\end{equation} 
The whole mass term can then be written as
\begin{equation}
        \mathcal{L}_M = - \overline{\widehat{\Psi}}_L M \widehat\Psi_R+ h.c.\, ,
\end{equation}
with
\begin{equation}
        M\equiv
\left( \begin{array}[!h]{cccccc}
        m_f^{(0,0)} & 0                         & m_f^{(0,1)} & 0                       & m_f^{(0,2)}& \dots\\
        m_f^{(1,0)} & M_{Q,1} & m_f^{(1,1)} & 0                         & m_f^{(1,2)}& \dots\\
        0                                       & 0                     & M_{q,1}                       & 0                     & 0                                      & \dots\\
        m_f^{(2,0)} & 0                         & m_f^{(2,1)} & M_{Q,2} & m_f^{(2,2)}& \dots\\
        0                                               & 0                     & 0                                             & 0                     & M_{q,2}                & \dots\\
        \vdots                  & \vdots  & \vdots      & \vdots  & \vdots     & \ddots
\end{array} \right)\, ,
\label{eqn:fermM}
\end{equation}
where $M_{\{Q,q\},n}$ denote the KK masses (before EWSB) \citep{delAguila:2000kb,Huber:2003tu}. The zeros are due to the fact that fields that are odd under the $Z_2$-parity vanish at the IR brane.

\subsection{Diagonalization of the Mass Matrix}

In order to diagonalize (\ref{eqn:fermM}) we use the important fact, that the entries of the mass matrix are not independent from each other. The Yukawa masses can be factorized in terms of profiles of the corresponding fermions, evaluated at the IR brane.
We define       
\begin{equation}
\begin{split}
        \al^L_m\equiv\sqrt{\epsilon^{-1} \la^{(5)}_{f}} f_m^{L(Q)}(\pi)\, ,\\
        \al^R_n\equiv\sqrt{\epsilon^{-1} \la^{(5)}_{f}} f_n^{R(q)}(\pi)\, ,
\end{split}
\end{equation} 
so that
\begin{equation}
        m_{f}^{(m,n)}=\frac{v}{\sqrt 2}\, \al^L_m\al^R_n.
\end{equation}
Now the mass matrix reads
\begin{equation}
        M=\frac 1{\sqrt 2}\left( \begin{array}[!h]{cccccc}
        v\al^L_0\al^R_0 & 0                       & v\al^L_0\al^R_1 & 0                     & v\al^L_0\al^R_2& \dots\\
        v\al^L_1\al^R_0 & \sqrt 2 M_{Q,1} & v\al^L_1\al^R_1 & 0                     & v\al^L_1\al^R_2& \dots\\
        0                                       & 0                     & \sqrt 2 M_{q,1}                       & 0                     & 0                                      & \dots\\
        v\al^L_2\al^R_0 & 0                       & v\al^L_2\al^R_1 & \sqrt 2 M_{Q,2} & v\al^L_2\al^R_2& \dots\\
        0                                               & 0                     & 0                                             & 0                     & \sqrt 2 M_{q,2}                & \dots\\
        \vdots                  & \vdots  & \vdots      & \vdots  & \vdots     & \ddots
\end{array} \right)\, .
\end{equation}
Deriving the characteristic polynomial of $M M^\dagger$ leads again to a transcendental equation for the (squared) mass eigenvalues of the fermion modes $M_f^{(m)2} \equiv \la^f_m$. We find
\begin{equation}
        \la^f_m - {\la^f_m}^2\, \frac{v^2}{2} \sum_{j,k=0}\frac{{[\al^L_j]}^2{[\al^R_k]}^2}{(M_{Q,j}^2-\la^f_m)(M_{q,k}^2-\la^f_m)}=0\, ,
        \label{eqn:fermtrans}
\end{equation}
where we have assumed $\la^{(5)}_{f}$ to be real-valued. 
Note that $M_{Q,0}=M_{q,0}=0$.

\subsection{The Spectrum}

Since the transcendental equation (\ref{eqn:fermtrans}) can not be solved analytically, we perform again an expansion in powers of $v^2/M_{KK}^2$ and obtain
\begin{equation}
        M_f^{(0)\,2} =m_f^{(0,0)\,2}\left[1-\frac{v^2}{2} \left({[\al^R_0]}^2\sum_{n=1}\frac{{[\al^L_n]}^2}{M^2_{Q,n}} + {[\al^L_0]}^2\sum_{n=1}\frac{{[\al^R_n]}^2}{M^2_{q,n}} \right) +\mathcal{O}\Big{(}\frac{v^4}{M_{KK}^4}\Big{)}\right]\, .
\label{0masses}
\end{equation}
One can see that the physical zero mode mass is lowered compared to the leading term in the expansion. For the KK modes we get 
\begin{align}
\label{sdmasses}
        M_f^{(m)2} = \begin{cases} 
M^2_{Q,\frac{m+1}{2}}\left[\displaystyle 1+\frac{v^2}2 {[\al^L_{\frac{m+1}2}]}^2 \sum_{n=0}\frac{{[\al^R_n]}^2}{{M^2_{Q,\frac{m+1}{2}}}-{M^2_{q,n}}}+\mathcal{O}\Big{(}\frac{v^4}{M_{KK}^4}\Big{)}\right]\, , & \text{for odd }m\\[2mm]
M^2_{q,\frac m 2}\left[\displaystyle 1+\frac{v^2}2 {[\al^R_{\frac m 2}]}^2 \sum_{n=0}\frac{{[\al^L_n]}^2}{{M^2_{q,\frac m 2}}-{M^2_{Q,n}}}+\mathcal{O}\Big{(}\frac{v^4}{M_{KK}^4}\Big{)}\right]\, , & \text{for even }m.\end{cases}
\end{align}
Here $m$ labels all diagonal entries of the (diagonalized) mass matrix, and no longer charged or neutral fields separately. One observes, that to first order in $v^2/M_{KK}^2$ just neutral states of other KK levels contribute to the mass corrections of the charged fermion and vice versa.
\begin{figure}[t!]
        \begin{center}
  \includegraphics[width=7.25cm]{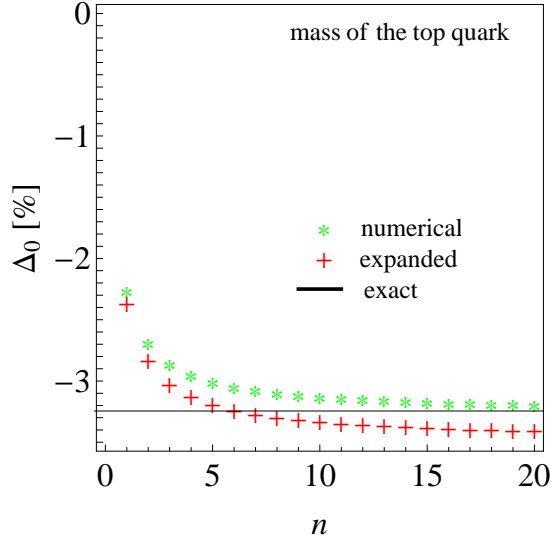}
        \end{center}
        \caption{Dependence of the fermion zero mode mass on the truncation.}
\label{fig:fermcon0}
\end{figure}

The explicit form of the fermion profiles that enter the above equations was calculated in \citep{Gherghetta:2000qt,Grossman:1999ra,Casagrande:2008hr}. Defining $c_{Q,q}=\pm m_{Q,q}/k$, where $m_{Q,q}$ is the bulk mass term of the corresponding fermion, the zero mode solutions read
\begin{equation}
\label{eqn:zeromode}
        f_0^{L,R}(\phi)=N_{L,R}\sqrt{\frac{L\epsilon}{\pi}}\, t^{c_{Q,q}}\, ,\quad N_{L,R}^2=\frac{1/2 + c_{Q,q}}{1-\epsilon^{1+ 2c_{Q,q}}}.     
\end{equation}
For the excited states ($n>0$), we define $x_{Q,q}^{(n)}=M_{\{Q,q\},n}/M_{KK}$ and the solutions are given by
\begin{equation}
\begin{split}
   f_n^{\{L(Q),R(q)\}}(\phi)
   &= {\cal N}(c_{\{Q,q\}},x_{\{Q,q\}}^{(n)})\,\sqrt{\frac{L\epsilon t}{\pi}}\,
    f^+(t,c_{\{Q,q\}},x_{\{Q,q\}}^{(n)}) \,, \\
    f_n^{\{R(Q),L(q)\}}(\phi)
   &= \pm {\cal N}(c_{\{Q,q\}},x_{\{Q,q\}}^{(n)})\,\mbox{sgn}(\phi)\,
    \sqrt{\frac{L\epsilon t}{\pi}}\,f^-(t,c_{\{Q,q\}},x_{\{Q,q\}}^{(n)}) \,,
\end{split}
\end{equation}
where
\begin{equation}
\begin{split}
 f^\pm(t,c,x) 
   = J_{-\frac12-c}(x\, \epsilon)\,J_{\mp\frac12+c}(x\,  t) 
   \pm J_{\frac12+c}(x\, \epsilon)\,J_{\pm\frac12-c}(x\,  t),\\
   {\cal N}^{-2}(c,x) 
   = \left[ f_n^+(1,c,x) \right]^2 - \epsilon^2 \left[ f_n^+(\epsilon,c,x) \right]^2 \,.
\end{split}
\end{equation}
The (unperturbed) KK masses $M_{\{Q,q\},n}$ are now obtained from the BC at the IR brane
\begin{equation}
   f^-(1,c_{\{Q,q\}},x_{\{Q,q\}}^{(n)}) = 0.
\end{equation}
In the exact approach the masses $M_n$ of the physical fermions are determined by the equation \citep{Casagrande:2008hr}
\begin{equation}
\label{eqn:feex}
   1-{\la^{(5)}_{t}}^2 \frac{L^2\, v^2}{8\pi^2\, M_{KK}^2}\, \frac{f^+(1,c_{q},x^{(n)})f^+(1,c_{Q},x^{(n)})}{f^-(1,c_{q},x^{(n)})f^-(1,c_{Q},x^{(n)})} = 0,
\end{equation}
where the index $n$ labels all mass eigenstates.
\begin{figure}[t!]
        \begin{center}
  \includegraphics[width=7.25cm]{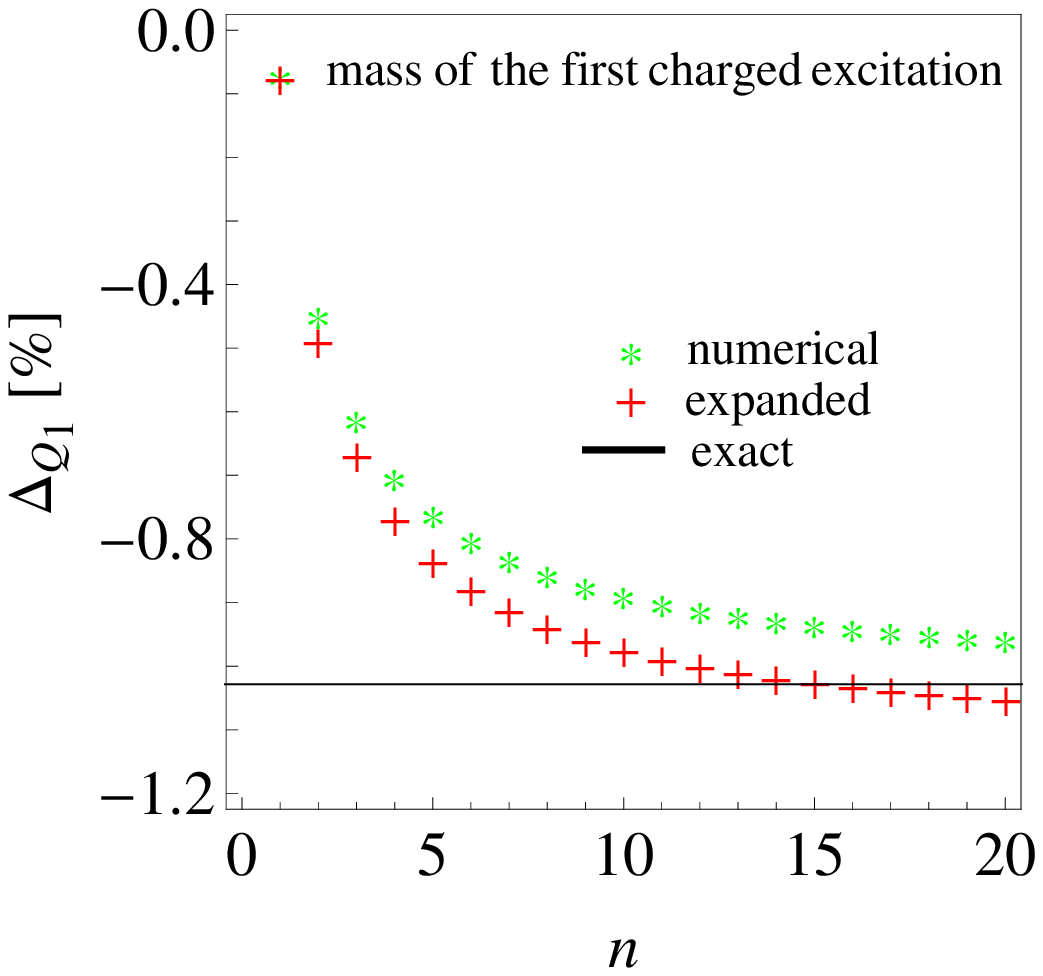}\qquad \includegraphics[width=7cm]{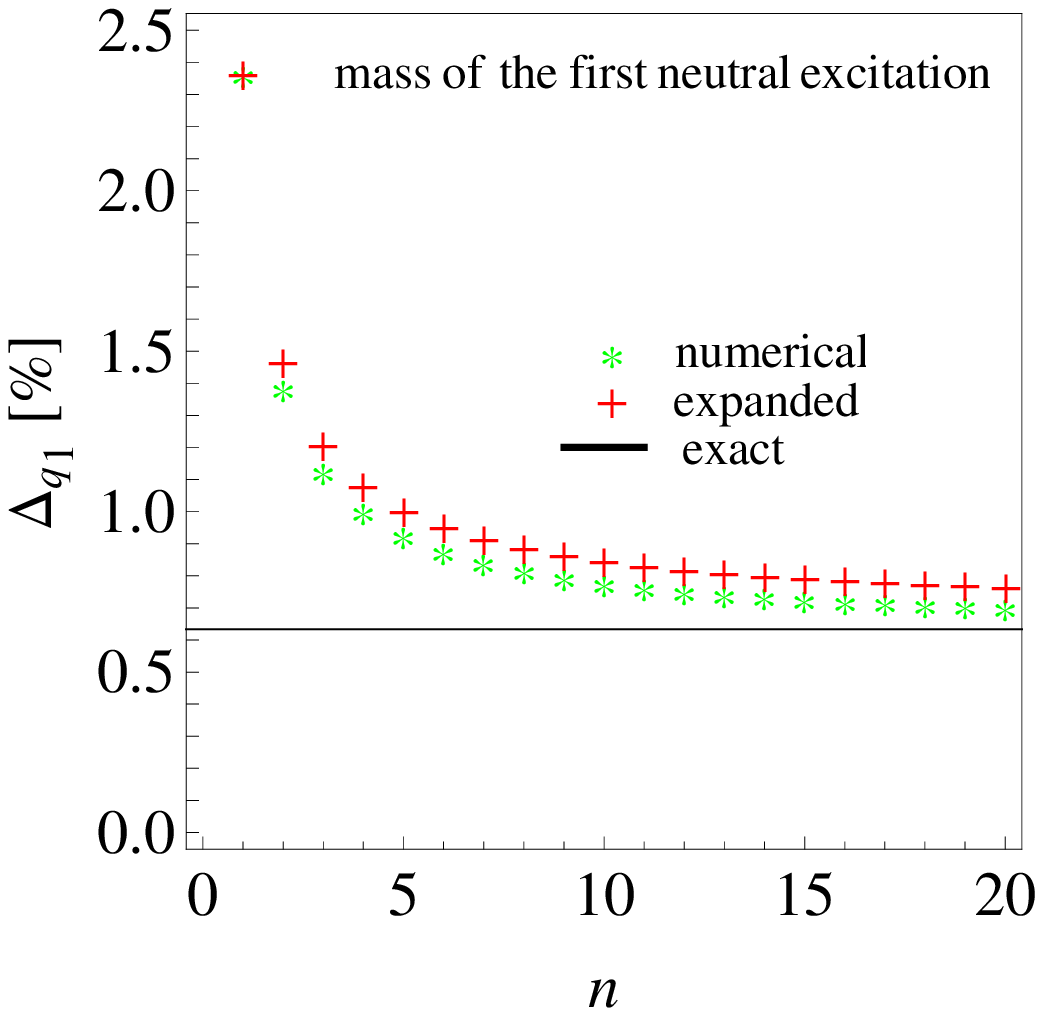}
        \end{center}
        \caption{Dependence of the masses of the first KK level on the truncation.}
\label{fig:fermcon1}
\end{figure}

In Figures \ref{fig:fermcon0} and \ref{fig:fermcon1} we compare the masses obtained from the different approaches, demonstrating the dependence on the truncation of the infinite sums in the corresponding expressions. For illustration we consider the top quark and take the values
\begin{equation}
\label{eqn:fermpara} 
c_Q = -0.473\, , \quad c_t = 0.339 \, , \quad \la^{(5)}_{t} = 0.422
\end{equation}
from \citep{Casagrande:2008hr} as input parameters. We plot again the relative deviations of the perturbed masses from the bare ones. These would be $m_f^{(0,0)}=140$\, GeV for the zero mode, $M_{Q,1}=3.690$\, TeV for the first charged and $M_{q,1}=5.419$\, TeV for the first neutral excitation for our choice of parameters. The green (light gray) stars correspond to a numerical solution of (\ref{eqn:fermtrans}), truncating the infinite sums after $n$ modes.  This way, $n$ charged and $n$ neutral modes are taken into account.
The red (dark gray) crosses represent the masses we get from the expansions (\ref{0masses}) and (\ref{sdmasses}), truncated in the same way. The black lines show the exact results obtained by using the formalism of \citep{Casagrande:2008hr} (see (\ref{eqn:feex})). One observes that the perturbative approach works quite well. It is interesting to note that taking into account just the mixing with the first KK level, which includes the first excitation of the charged fermion as well as that of the neutral one, already accounts for about $70 \%$ of the total shift in the zero mode mass. For the excitations this is not the case, but nevertheless the convergence is reasonable. Note that also for the fermions the expansion in $\mathcal{O}(v^2/M_{KK}^2)$ works well.
The eigenvectors, which build the entries of the diagonalization matrices $U$ and $V$ with $U^\dagger M V=\tilde M$, where $\tilde M$ is the diagonalized mass matrix, are given in Appendix \ref{app:fermev}.

For $N$ generations of fermions the entries in the mass matrix (\ref{eqn:fermM}) are replaced by $N \times N$ matrices. Therefore the diagonalization and derivation of the eigenvalues presented in this section does not work any longer. One could perform the diagonalization, using an expansion in $v^2/M_{KK}^2$ from the very beginning (see also \citep{delAguila:2000kb}). However, this leads to quite complicated, implicit expressions for the entries of the diagonalization matrices and the method presented in \citep{Casagrande:2008hr} seems more suitable for the $N$ fermion case.

\section{Conclusions}

\label{sec:Conclusions}

In this paper, we have worked out the effective 4D theory for the spontaneously broken $SU(2)_L \times U(1)_Y$ gauge symmetry in the Randall-Sundrum geometry, applying the widely used perturbative approach. We have performed the gauge fixing in a covariant $R_\xi$ gauge and pointed out its relevance for loop calculations in the decomposed theory. The emerging mass matrices have been diagonalized in closed form and a comparison of the spectrum with the results obtained by the exact method of \citep{Casagrande:2008hr} has been presented, emphasizing that already a low truncation of the KK tower leads to good numerical agreement. Then we have applied the perturbative approach to the fermion sector, where we restricted the considerations to the case of a single fermion. In summary, we have validated the applicability of the perturbative approach. For calculations incorporating $N$ generations however, the exact method seems to be more suitable.

\subsection*{Acknowledgments}

We are grateful to Sandro Casagrande, Uli Haisch and Matthias Neubert for useful discussions and encouragement.

\newpage
\begin{appendix}
\begin{landscape}

\section{Eigenvectors in the Fermion Sector}
\label{app:fermev}
\renewcommand{\theequation}{A\arabic{equation}}
\setcounter{equation}{0}

The relations between mass eigenstates (denoted by a tilde) and interaction eigenstates for the fermions read:
\begin{equation}
\begin{split}
	\widehat{\tilde\Psi}^{(n)}_L=& \sum_m U^{\dagger(n,m)} \widehat\Psi^{(m)}_L\\
	=& \sum_{m=1} \frac{1}{N^U_n}\left\{\psi^{L(Q)}_0 - \left(\frac{\al^L_m \la^f_n}{\al^L_0 (M^2_{Q,m}-\la^f_n)}\right) \psi^{L(Q)}_m + \left(\frac{\al^R_m M_{q,m}}{\al^L_0 v (M^2_{q,m}-\la^f_n)\left(\sum_{k=0}\frac{\al^{R2}_k}{M^2_{q,k}-\la^f_n}\right)} \right) \psi^{L(q)}_m \right\},\\
	N^U_n=&\sqrt{1+\sum_{m=1}\left(\frac{\al^L_m \la^f_n}{\al^L_0 (M^2_{Q,m}-\la^f_n)}\right)^2+\sum_{m=1}\left(\frac{\al^R_m M_{q,m}}{\al^L_0 v (M^2_{q,m}-\la^f_n)\left(\sum_{k=0}\frac{\al^{R2}_k}{M^2_{q,k}-\la^f_n}\right)} \right)^2},
\end{split}
\end{equation}
and
\begin{equation}
\begin{split}
	\widehat{\tilde\Psi}^{(n)}_R=&\sum_m V^{\dagger(n,m)} \widehat\Psi^{(m)}_R\\
	=&\sum_{m=1} \frac{1}{N^V_n}\left\{\psi^{R(q)}_0 +\left(\frac{\al^L_m M_{Q,m}}{\al^R_0 v(M^2_{Q,m}-\la^f_n)\left(\sum_{k=0}\frac{\al^{L2}_k}{M^2_{Q,k}-\la^f_n} \right)} \right) \psi^{R(Q)}_m - \left(\frac{\al^R_m \la^f_n}{\al^R_0 (M^2_{q,m}-\la^f_n)}\right) 
 \psi^{R(q)}_m \right\},\\
	N^V_n=&\sqrt{1+\sum_{m=1}\left(\frac{\al^L_m M_{Q,m}}{\al^R_0 v(M^2_{Q,m}-\la^f_n)\left(\sum_{k=0}\frac{\al^{L2}_k}{M^2_{Q,k}-\la^f_n} \right)} \right)^2+\sum_{m=1}\left(\frac{\al^R_m \la^f_n}{\al^R_0 (M^2_{q,m}-\la^f_n)}\right)^2}.
\end{split}
\end{equation}
Looking at these relations one observes a mixing of neutral fermions with charged ones, so that for the SM gauge group there could be right-handed couplings of zero mode fermions to the $W$ boson \citep{Huber:2003tu, delAguila:2000kb,Hewett:2002fe,delAguila:2000rc}.
\end{landscape}
\end{appendix}

\end{document}